\documentclass{aa}
\def\msol{{M}_{\odot}}

\def\grsim{\,\lower 1mm \hbox{\ueber{\sim}{>}}\,}
\def\lesssim{\,\lower 1mm \hbox{\ueber{\sim}{<}}\,}
\def\ueber#1#2{{\setbox0=\hbox{$#1$}%
  \setbox1=\hbox to\wd0{\hss$ #2$\hss}%
  \offinterlineskip
  \vbox{\box1\box0}}{}}
\input psfig

\usepackage{aas_macros}
\usepackage[longnamesfirst]{natbib}
\usepackage{graphics}
\bibpunct[;]{(}{)}{;}{a}{}{,}

\begin{document}

%   \thesaurus{03     % A&A Section 3: extragalactic           %JO
%              ( 11.03.1; %{\bf Galaxies: clusters: general    %JO
%                11.09.3; %intergalactic medium                %JO
%%                12.03.3; %Cosmology: observations             %JO
%                12.03.4; %Cosmology: theory                   %JO
%                12.04.1; %dark matter                         %JO
%                13.25.2)} % X-rays: galaxies                  %JO
%
   \title{Discovery of depressions in the X-ray emission of the distant galaxy
   cluster RBS797 in a CHANDRA observation}

%   \subtitle{}

   \author{S. Schindler\inst{1},
           A. Castillo-Morales\inst{1},
           E. De Filippis\inst{1},
           A. Schwope\inst{2},
           J. Wambsganss\inst{3}
%          \and ...\inst{2}
%          C. Ptolemy\inst{2}\fnmsep\thanks{Just to show the usage
%          of the elements in the author field}
          }

   \offprints{S. Schindler}

   \institute{Astrophysics Research Institute,              %JO
               Liverpool John Moores University,             %JO
	       Twelve Quays House,                           %JO
               Birkenhead CH41 1LD,                          %JO
               United Kingdom;
\and 
Astrophysikalisches Institut Potsdam,
               An der Sternwarte 16,
               14482 Potsdam,
               Germany;
\and
Universit\"at Potsdam, Institut f\"ur Physik, Am Neuen Palais 10, 14469
               Potsdam, Germany}    

%         \and
%             University of Alexandria, Department of Geography\\
%             email: c.ptolemy@hipparch.uheaven.space
%             \thanks{The university of heaven temporarily does not
%                     accept e-mails}

   \date{}

\abstract{
We present CHANDRA observations of the X-ray luminous, distant 
galaxy cluster RBS797 at $z=0.35$. In the
   central region the X-ray emission shows two pronounced X-ray
   minima, which are located opposite to each other with respect to
   the cluster centre. These depressions suggest an interaction
between the central radio galaxy and the intra-cluster medium, which
would be the first detection in such a distant cluster. The minima are 
symmetric relative to the cluster centre
and very deep compared to similar features found in a few 
other nearby clusters. A
spectral and morphological analysis of the overall cluster emission
shows that RBS797 is a hot cluster ($T=7.7^{+1.2}_{-1.0}$keV) with a total
mass of $M_{\rm tot}(r_{500})=6.5^{+1.6}_{-1.2}
\times10^{14}\msol$.
}
\authorrunning {S. Schindler  et al.}
\titlerunning {CHANDRA observation of RBS797}
   \maketitle

%
%________________________________________________________________

\section{Introduction}

Interaction between the intra-cluster gas and radio sources in the
centre of galaxy 
clusters has been found in several nearby clusters with an
anticorrelation of X-ray emission and radio emission, e.g. in the 
Perseus cluster \cite[]{fabian00},
in Hydra-A \cite[]{mcnamara00},
and in Abell clusters A4059 \cite[]{huang98},
A2597 \cite[]{mcnamara01},
A2052  \cite[]{sarazin01},
A2199  \cite[]{fabian01},
and A2634 \cite[]{sas96}.
Such an interaction of two different components is particularly
interesting because                             one can infer physical
conditions and processes in the cluster centre, e.g. heating and
cooling effects, magnetic fields or energies of relativistic
particles, from the pressure  balance.
           The observations suggest that the cavities in the X-ray gas
rise outwards by buoyancy \cite[]{mcnamara00,churazov00} while a model
by \cite{heinz98} predicts shocked gas around the
cavities which is 
not observed. Models by \cite{reynolds01} and \cite{david00} suggest 
a weak shock
regime with radio lobes moving at the local sound speed.

We present here recent CHANDRA observations of the galaxy
cluster RBS797.
They reveal clear X-ray minima in the cluster centre. We hypothesise
that these minima are caused by interaction
between the radio source embedded in the
central cluster galaxy and the intra-cluster medium. 
RBS797 would be the first {\em distant} cluster in which evidence for 
such interaction has been found. 
Only the high spatial resolution of CHANDRA makes it 
possible to observe such details in distant clusters. 
This X-ray observation  is part of a 
broader programme to search for strong gravitational lensing in
X-ray selected, X-ray luminous clusters.  
The goal is to take both deep optical images of these
clusters and high resolution X-ray exposures. 
In this way  we can use the clusters
for arc statistics as well as for
the determination and comparison of cluster masses
estimated by X-ray observations and by gravitational lensing. 
Throughout this paper we use $\rm{H}_0 = 50$ km/s/Mpc and $\rm{q}_0=0.5$.

%__________________________________________________________________

\section{Cluster identification}

The X-ray source RXS J094713.2+762317 (or RBS797), 
found in the ROSAT All-Sky 
Survey (RASS),
was observed optically in the course of the      ROSAT Bright
Survey (RBS), aiming at optical identification of all bright
(count rate $> 0.2$\,cts\,s$^{-1}$), high-galactic latitude ($|b| >
30\degr$) X-ray 
sources detected in the RASS \cite[]{fischer98,schwope00}.
%(Fischer et al.~1998, Schwope et al.~2000).
A %single 
spectrum of the central cluster galaxy at RA(2000) = 09 47 12.5, DEC(2000) =
+76 23 14, taken with the SAO 6m telescope shows a strong O[II]$\lambda$3727 
emission line, and weak, narrow H$\beta$ and O[III]$\lambda\lambda$4959/5007 
emission lines at a redshift of $z = 0.354$. 
The emission lines can be due to a cooling flow or
due to nuclear activity of the central 
cluster galaxy.

%__________________________________________________________________

\section{X-ray observation of RBS797 with CHANDRA}

RBS797 was observed on October 20$^{th}$, 2000  with the
CHANDRA Advanced CCD Imaging Spectrometer (ACIS) I
detector for a total exposure time of 13.3 ksec. No time was lost due
to flares. For the data analysis
we used the standard set of event grades and applied no a
priori filtering. The energy range used for the images is 0.5-7~keV, 
for the spectra it is  given in Table~1. 
The response matrices from the CHANDRA Calibration Database package
(CALDB) version 2.0 are used.
 
\begin{figure}
\hskip 0.6cm
\psfig{figure=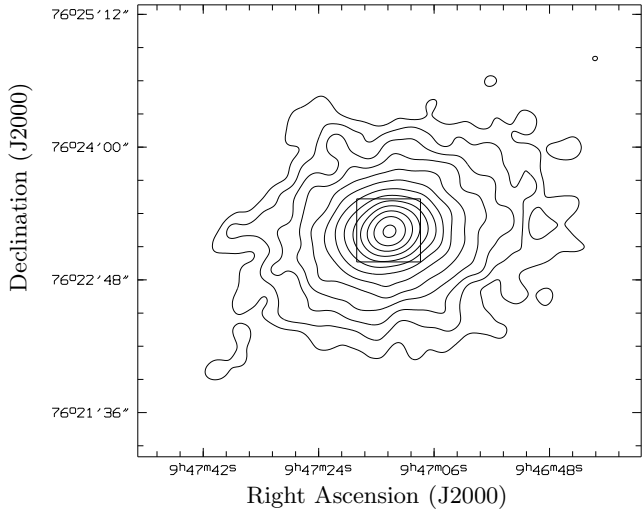,width=7.9cm,clip=}
\put(-240.,70.){\rotatebox{90}{Declination~(J2000)}}
\put(-150.,-10.){Right Ascension~(J2000)} 
\caption[]{CHANDRA image of the cluster RBS797. The cluster is relatively
regular with an ellipticity of 1.3-1.4 in E-W direction.
The contours are
logarithmically spaced with 5 contours per decade
and the highest corresponding to 3.9 cts/s/arcmin$^2$. 
The image is smoothed with a Gaussian of
$\sigma=5$ arcsec. The size of the image is 4.7 arcmin on the side.
The central square marks the region shown with higher resolution 
in Fig.~\ref{fig:cont_arrow}.
}
\label{fig:cont_tot}
\end{figure}

Figure~\ref{fig:cont_tot} shows the X-ray image of the cluster RBS797 on
a scale of 4.7~arcmin ($\equiv 1.7$~Mpc). The cluster emission is
elliptical with 
axis ratios of major to minor axis varying slightly from 1.3 at a
radius of 0.26 arcmin to 1.4 at a radius of
1.7 arcmin. The centre and the position angle ($\approx -70^{\circ}$, N
over E) of the ellipses are almost the same over the entire radius range. 

\begin{figure}
\psfig{figure=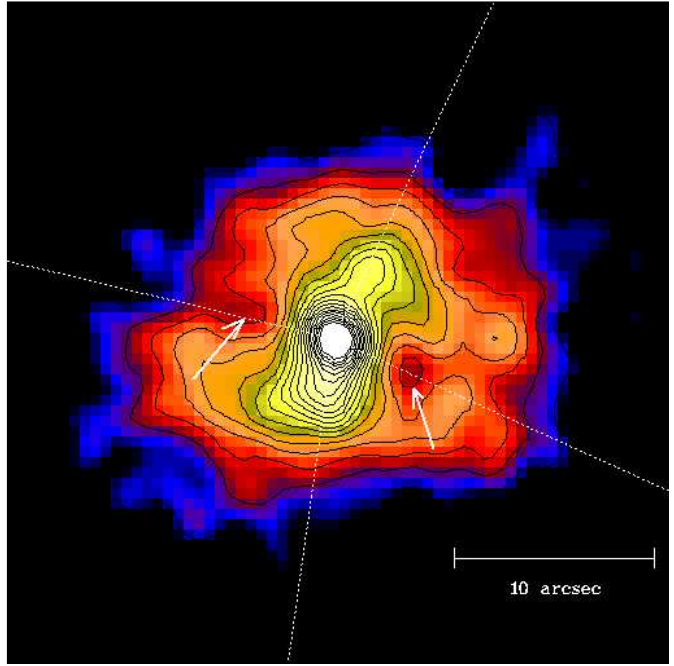,width=8.8cm,clip=} 
\caption[]{X-ray image of the central (32 arcsec)$^2$ of the
cluster RBS797. There are clear minima in the X-ray emission in ENE and WSW
direction at about 5 arcseconds from the cluster centre (see arrows). 
An excess of emission is found in perpendicular directions. The dotted
lines mark the traces shown in Fig.~\ref{fig:trace_bw}.
The contours have a linear spacing of
$0.7$ cts/s/arcmin$^2$ with the highest being
$14$ cts/s/arcmin$^2$. The image is smoothed with a Gaussian of
$\sigma=0.75$ arcsec.
}
\label{fig:cont_arrow}
\end{figure}

\begin{figure}
\psfig{figure=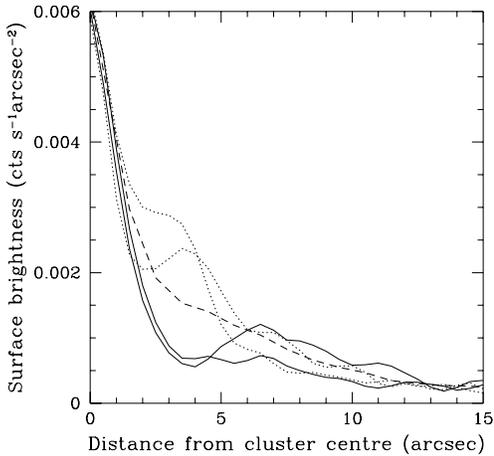,width=6.5cm,clip=} 
\caption[]
{Dependence of the surface brightness on direction: 
the X-ray emission from the centre in direction
of the minima (solid lines, position angles $77^{\circ}$ and
$243^{\circ}$, N over E)  
and in direction of the maxima
(dotted lines, position angles $171^{\circ}$ and
$337^{\circ}$). 
There is an X-ray deficit of a factor of 3 - 4
in the ``holes'' compared to the perpendicular directions.  
The dashed line shows an average profile integrated over all angles.
A surface
brightness of 0.001 cts/s/arcsec$^2$ corresponds to 3 counts per ACIS
pixel (0.5 arcsec on the side).
}
\label{fig:trace_bw}
\end{figure}

The innermost (32 arcsec)$^2$ 
region of the galaxy cluster RBS797 is shown in
Figure~\ref{fig:cont_arrow}. Two emission minima are obvious in ENE and WSW 
direction at distances of about 3-5 arcseconds from the cluster
centre. The minima 
are opposite to each other with respect to the cluster
centre. In perpendicular directions
excess X-ray emission is visible. At radii of 4 arcsec there is a
factor of 3-4 between the X-ray emissions from the different regions
(see Fig.~\ref{fig:trace_bw}). The sizes  of the X-ray depressions are
about 20 - 30 kpc.
It is likely -- but largely a hypothesis up to now -- that the gas has been
pushed from the 
low X-ray emission regions to the high X-ray emission regions by the
pressure of relativistic particles in radio lobes, as has been
observed in a number of nearby clusters. Although a good hardness
ratio map cannot be derived from the data due to the limited number of
photons, we find that the hardness ratio in the depressions is
not harder than than the overall
cluster emission, which excludes
absorption or hot gas bubbles as explanations for the minima.
The depressions are very significant and deep, and also very symmetric
compared to other clusters. Therefore this cluster appears to be an ideal
object to investigate the interaction of jets and the
intra-cluster medium. We will study the interaction 
in detail with radio follow-up observations.

%__________________________________________________________________

\section{Spectral analysis, luminosity and mass determination}

The CHANDRA data can be used to derive the temperature, 
metallicity, luminosity and mass of RBS797.  
For the spectral fits the
energy range is restricted to $0.5-10$~keV and $2-10$~keV, respectively,
because the calibration of the low energy channels is still
uncertain. 
%Furthermore, numerical tests showed that
%cluster temperatures can be 
%contaminated by material falling into the cluster
%\cite[]{mathiesen01}, when taking into account the low energy range.
We treat the central
point-like source  (radius $\lesssim 2$ arcseconds) separately, 
because this
source is probably an AGN, which would distort the thermal spectrum of
the intra-cluster gas.

The results of the spectral analysis are summarised in
Table~\ref{tab:spectral}. 
The overall emission is well fit by a thermal bremsstrahlung model with a
temperature of 7.7 keV and a metallicity around 0.26 in
solar units when the hydrogen column density is fixed to
the Galactic value \cite[]{dickey90}. 
%The results do not change when
%the energy binning is varied; for Table~\ref{tab:spectral} a binning of 20
%counts/bin is used.
We find somewhat higher temperatures for fits in the 0.5-10 keV band,
compared to the results of the 2-10 keV band, probably indicating 
problems with the calibration of the low energy channels.
Attempting to see radial variations in the fit parameters, we extract
photons from three different annuli. 
%(here the binning is 25 counts per spectral bin). 
There is a trend of increasing
temperature and decreasing metallicity with radius, which, however,
is not statistically 
significant due to the limited number of photons in this pointing.

The best-fit value for the redshift ($z = 0.39\pm0.02$) 
is slightly higher than the
optically determined redshift ($z_{\rm opt} =  0.354$), but they
still agree with each other within $2\sigma$. If we fix
the redshift to the optical value, the resulting metallicity is lower because
the location of the Fe lines does not correspond well with the bump in
the observed X-ray spectrum.

Although the ``arms'' -- the regions of higher emission
surrounding the minima -- contain
only 1387 source counts, we attempt a spectral fit, because the
temperature measurement is important for distinguishing different
models. We find a slightly lower temperature
(4.4~keV, see Table~\ref{tab:spectral}) 
than for the rest of the cluster, suggesting that the
``arms'' are probably not shocked regions. This would 
rule out the model by \cite{heinz98}.
Similar results have been found in other clusters 
\cite[]{fabian00,mcnamara00}.

\begin{table*}[t]
\caption[]{Results of the spectral fits to the X-ray emission of
the galaxy cluster RBS797. Column 1 shows the radii in
arcsec of
the region, from which the photons have been extracted. Column 2
lists the model used for the fit:  MeKaL =
\cite{meka93}, \cite{liedahl95}, PL = power law. 
Column 3 lists the energy range. In
columns 4-7 the fit parameters are given: hydrogen column density in
$10^{20}$ cm$^{-2}$, temperature in keV (for power law fits it is the
power law index), metallicity in solar units,
redshift. The errors are 90\%
confidence levels. Columns 8 and 9 are the
degrees of freedom of the fit and the reduced $\chi^2$.}
\begin{center}
\begin{tabular}{|c|c|c|c|c|c|c|c|c|}
\hline
1 & 2 & 3 & 4 & 5 &  6 & 7 & 8 & 9 \cr 
\hline 
radius    & model & energy & $n_H$ in & temp. & metallicity Z & redshift & d.o.f.  & $\chi^2$/d.o.f. \cr
in arcsec    &       & range in keV  & 10$^{20}$ cm$^{-2}$  & kT in keV    & in solar units       & z        &         &                 \cr
\hline
$2-80$ & MeKaL & $0.5-10$       & 2.22(fixed) & $7.7^{+1.2}_{-1.0}$&   
$0.26\pm0.10$         & $0.39\pm0.02$           & 272 & 1.2     \cr
$2-80$ & MeKaL & $0.5-10$       & $6.1^{+1.5}_{-1.4}$ & $6.3\pm0.6$&   
$0.26\pm0.09$         & $0.38\pm0.02$           & 271 & 1.1     \cr
$2-80$ & MeKaL & $2-10$       & 2.22(fixed) & $8.1^{+1.6}_{-1.5}$&   
$0.25\pm0.12$         & $0.39\pm0.02$           & 109 & 0.9     \cr
$2-80$ & MeKaL & $2-10$       & 2.22(fixed) & $7.4^{+1.7}_{-1.1}$&   
$0.17\pm0.11$         &  0.354(fixed)            & 91  & 1.0     \cr
%$2-80$ & RS    & $0.5-10$       & 2.22(fixed) & $7.7^{+0.8}_{-0.6}$ &
%$0.23\pm 0.1$  & $0.39^{+0.02}_{-0.03}$          &  273&  1.2    \cr
%$2-80$ & RS    & $2-10$       & 2.22(fixed) & $8.0^{+1.7}_{-1.3}$ &
%$0.22^{+0.11}_{-0.10}$  & $0.39^{+0.02}_{-0.03}$& 110 &   0.9   \cr
$2-10$ & MeKaL & $0.5-10$       & 2.22(fixed) & $5.7^{+0.7}_{-0.5}$&   
$0.38^{+0.17}_{-0.16}$     &  0.39(fixed)            & 117 & 1.6     \cr
$10-30$ & MeKaL & $0.5-10$       & 2.22(fixed) & $8.4^{+1.2}_{-1.0}$&   
$0.25\pm0.18$     &  0.39(fixed)            & 139 & 1.0     \cr
$30-80$ & MeKaL & $0.5-10$       & 2.22(fixed) & $11.7^{+4.3}_{-2.5}$&   
$0.20^{+0.35}_{-0.20}$    &  0.39(fixed)            & 95  & 1.2     \cr
arms & MeKaL & $0.5-10$       & 2.22(fixed) & $4.4^{+0.7}_{-0.6}$&   
0.26(fixed)    &  0.39(fixed)            & 52  &  1.2    \cr
%$0-2$ & MeKaL & $0.5-10$       & 2.22(fixed) & $80$&   
%0.27$    &  0.39(fixed)            & 18  &  1.0    \cr
$0-2$ & PL & $0.5-10$       & 2.22(fixed) & $1.2\pm0.2$&   
 -    &  0.39(fixed)            &  20 &  0.9    \cr
$0-2$ & 0.59$\times$MeKal & $0.5-10$       & 2.22(fixed)
&$7.7$(fixed)    &
0.26(fixed)  &  0.39(fixed)            &  19 &   0.9   \cr
  & + 0.41$\times$PL &        & 
&$1.1\pm0.2$    
  &            &   &  &   \cr
\hline
\end{tabular}
\end{center}
\label{tab:spectral}
\end{table*}

In the centre of the cluster is a point-like source, which has a
source count rate of about 0.026 cts/s and a flux
$F_{\rm centre}$(2-10keV)$=2.2\times 10^{-13}$ erg/s/cm$^2$.  
The spectrum of this source is very flat
and cannot be fitted with a thermal spectrum of reasonable
temperature. A power law spectrum, which is an indication for an AGN,
(and also a sum of a  power law and
a thermal spectrum) 
yields a good fit with a very low 
index of $1.2\pm{0.2}$. 

The NRAO VLA Sky Survey (NVSS) 
lists a radio source at the central cluster position with a
peak flux of 0.02 Jy/beam \cite[]{condon98}
which is another indication for an active galaxy in the centre
of the cluster. A complementary 
explanation for the central X-ray peak would be a
cooling flow as the central cooling time is $\approx 10^9$~yr.

The X-ray emission of the cluster
can be traced out to a radius of 4.1~arcmin ($\equiv 1.5$~Mpc). 
Within this radius there are 11000 source counts. For the
above mentioned parameters, this corresponds to a flux of 
$f_X$(0.5-7.0 keV)$ = 7.1\times10^{-12}$erg/s/cm$^2$ and 
$f_X({\rm bol}) = 1.1\times10^{-11}$erg/s/cm$^2$ and for a redshift of
$z=0.354$ to a luminosity of
$L_X$(0.5-7.0 keV)$ = 4.0\times 10^{45}$erg/s and $L_X({\rm bol}) = 6.7 \times
10^{45}$erg/s. Only 3\% of the total
emission comes from the central source.

\begin{figure}
\psfig{figure=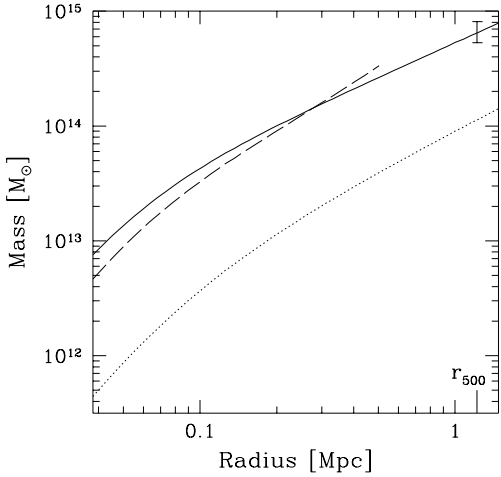,width=6.6cm,clip=} 
\caption[]
{Integrated mass versus radius: total gravitational mass (solid line)
and gas mass (dotted line). A typical error bar is shown at $r_{500}$.
The total mass profile taking into account a temperature gradient
(dashed line) is shown up to the radius to which the temperature
gradient can be determined.
}
\label{fig:mass}
\end{figure}

The X-ray surface brightness profile of RBS797 (centred on the central
galaxy) 
can be fitted well
by a $\beta$~model \cite[]{cavaliere76}
when ignoring the central region. A fit excluding the central
6 arcsec radius yields a slope $\beta=0.63\pm0.01$, a core
radius $r_c=8.1(\pm0.6)$~arcsec [or $49(\pm4)$~kpc] and a central surface
brightness $S_0=6.6$~cts/s/arcmin$^2$. 
With the assumption of spherical symmetry this 
corresponds to a gas mass of
$M_{\rm gas}$(1Mpc)$=0.90(\pm0.07)\times10^{14}\msol$ and 
$M_{\rm gas}(r_{500})=1.13(\pm0.09)\times10^{14}\msol$ with
$r_{500}= 1.22(\pm0.08)$~Mpc.
Assuming hydrostatic equilibrium, the total mass
of the cluster can be estimated. For an isothermal cluster of
$T=7.7$~keV the total mass is
$M_{\rm tot}$(1Mpc)$=5.3^{+0.8}_{-0.7}\times10^{14}\msol$ and 
$M_{\rm tot}(r_{500})=6.5^{+1.6}_{-1.2}\times10^{14}\msol$ (see
Fig.~\ref{fig:mass}). 
This corresponds to a gas mass fraction
$f_{\rm gas}(r_{500})=0.17^{+0.06}_{-0.05}$ which is
in good agreement with the gas mass fractions found in samples of
nearby and distant clusters
\cite[]{ettori99,mohr99,sas99}.
Taking into account the temperature gradient changes the
results slightly (see Fig.~\ref{fig:mass}).

%__________________________________________________________________

\section{Summary}

The X-ray luminous cluster RBS797 reveals remarkable depressions in
the X-ray emission of the central region. The depressions are roughly
circular and
arranged symmetrically  with respect to the cluster centre. They 
have diameters
of 20 - 30 kpc and are very deep: a factor of 3-4 less emission than
in the other directions. The discovery of these minima shows the
ability of CHANDRA to investigate physical effects even in clusters as
distant as RBS797 (z=0.35). 

The spectral and morphological analysis of RBS797 shows that
the cluster
has a high temperature of $T=7.7$~keV, a metallicity
$Z=0.2-0.3$, and a high luminosity
$L_{X}({\rm bol})=6.7\times10^{45}$erg/s. The total mass within $r_{500}$ is 
$M_{\rm tot}(r_{500})=6.5\times10^{14}\msol$ with a gas mass fraction
$f_{\rm gas}(r_{500})=0.17$.
In follow-up radio observations the interaction between the
radio lobes and the intra-cluster gas will be studied in detail.

%__________________________________________________________________

\begin{acknowledgements} 
We thank Doug Burke and Valeri Hambaryan for help with the
data analysis and Phil James for carefully reading the manuscript. 
\end{acknowledgements}

\bibliography{De143}
\bibliographystyle{aabib99}

%\begin{thebibliography}{}
%
%   \bibitem[1966]{baker} Baker N., 1966,
%      in: Stellar Evolution,
%      eds.\ R. F. Stein, A. G. W. Cameron,
%      Plenum, New York, p.\ 333
%
%\end{thebibliography}

\end{document}